\documentclass[twocolumn,showkeys,showpacs,preprintnumbers,prd, superscriptaddress,nofootinbib,aps,10pt,longbibliography]{revtex4-1}
\usepackage{graphicx,epsf,bm,amsmath,amsfonts,amssymb,epstopdf,natbib,color,verbatim,multirow,bm,mathtools,mathrsfs,braket,bbold,xcolor}
\usepackage{hyperref}
\usepackage[normalem]{ulem}
\hypersetup{colorlinks=true,urlcolor=blue,citecolor=blue,linkcolor=blue,menucolor=blue,anchorcolor=blue,filecolor=blue}
\date{\today}

\newcommand{\GeV}{\mathrm{GeV}}
\newcommand{\keV}{\mathrm{keV}}
\newcommand{\MeV}{\mathrm{MeV}}

\newcommand{\dd}{\mathrm{d}}
\newcommand{\qLZ}{\Delta q_{\mathrm{LZ}}}
\newcommand{\lprod}{\lambda_{\chi G}}

\begin{document}

\title{ALP-mediated inelastic dark matter and the LUX-ZEPLIN high-recoil candidate event LZ230616}

\author{Guan-Wen Yuan$^{*}$}
\affiliation{Department of Astronomy, School of Physical Sciences, University of Science and Technology of China, Hefei, Anhui 230026, China}
\affiliation{School of Astronomy and Space Science, University of Science and Technology of China, Hefei, Anhui 230026, China}

\author{Bo Zhang$^{\dagger}$}
\affiliation{Key Laboratory of Dark Matter and Space Astronomy, Purple Mountain Observatory, Chinese Academy of Sciences, Nanjing 210033, China}
\affiliation{School of Astronomy and Space Science, University of Science and Technology of China, Hefei, Anhui 230026, China}

\author{Wen-Yu Cao$^{\ddagger}$}
\affiliation{Department of Physics, The Chinese University of Hong Kong, Shatin, New Territories, Hong Kong}

\author{Lei Feng$^{\S}$}
\affiliation{Key Laboratory of Dark Matter and Space Astronomy, Purple Mountain Observatory, Chinese Academy of Sciences, Nanjing 210033, China}
\affiliation{School of Astronomy and Space Science, University of Science and Technology of China, Hefei, Anhui 230026, China}

\author{Ruizhi Yang$^{\P}$}
\affiliation{Department of Astronomy, School of Physical Sciences, University of Science and Technology of China, Hefei, Anhui 230026, China}
\affiliation{School of Astronomy and Space Science, University of Science and Technology of China, Hefei, Anhui 230026, China}

\begin{abstract}

The LUX-ZEPLIN (LZ) Collaboration has reported a high-energy candidate event LZ230616 with a reconstructed nuclear recoil energy $E_R=248\pm23_{\rm stat}\pm23_{\rm sys}~{\rm keV}$.
We investigate a possible interpretation in terms of inelastic scattering between two Majorana dark matter states mediated by an axionlike particle coupled to gluons. The positive mass splitting suppresses low-energy recoils, while the momentum dependence of the interaction reshapes the high-energy spectrum. 
We treat the dark-sector and gluonic couplings independently and retain the momentum-dependent nucleon form factors and xenon nuclear responses. Using an approximate single-event likelihood, we find that, for $m_a=0.3~{\rm GeV}$, a narrow spectrum near the candidate energy arises at $m_\chi\simeq0.35~{\rm TeV}$ and $\delta\simeq330~{\rm keV}$, although this configuration requires a large coupling product and is highly sensitive to the Galactic halo speed cutoff. 
Our analysis establishes the kinematic and coupling requirements for subsequent tests using the thermal relic abundance and laboratory constraints on the mediator.

\end{abstract}

\date{\today}

\maketitle

\section{Introduction}\label{sec:introduction}

The LUX-ZEPLIN (LZ) Collaboration recently extended its nuclear-recoil search to $E_R\simeq270\,\mathrm{keV}$ using an exposure of $2.84\,\mathrm{tonne\,yr}$ and reported a candidate event at $E_R=248\pm23_{\mathrm{stat}}\pm23_{\mathrm{sys}}\,\mathrm{keV}$ in a region of low expected background~\cite{LZ:2026axp}. The largest local significance among the tested interaction models was $3.4\sigma$, corresponding to a global significance of $2.6\sigma$. Although this observation does not establish a dark matter (DM) origin, its unusually high recoil energy motivates further investigation. The corresponding momentum transfer, $Q=\sqrt{2m_{\mathrm{Xe}}E_R}\simeq246\,\mathrm{MeV}$, exceeds that typically probed in conventional direct-detection searches. For momentum-independent elastic scattering, the available halo phase space and nuclear form factors generally suppress high-energy recoils. The LZ candidate therefore motivates interactions whose kinematics or matrix elements enhance the relative contribution of large recoil energies, as anticipated in nonrelativistic effective theories and previous high-recoil searches~\cite{Fan:2010gt,Fitzpatrick:2012ix,Anand:2013yka,XENON100:2017EFT,XENON1T:2024EFT,LZ:2024EFT}.

This observation builds on substantial progress in direct DM searches. Liquid-xenon experiments now achieve tonne-year exposures and provide leading constraints on DM--nucleon interactions~\cite{Balazs:2024Primer,Xia:2026DirectDM,Billard:2021uyg,LZ:2023First,PandaX:2021First,XENON:2023First,LZ:2025WIMP,PandaX:2025WIMP,XENON:2025WIMP}. Recent measurements and searches involving ${}^{8}\mathrm{B}$ solar neutrinos also highlight the increasing importance of neutrino backgrounds~\cite{PandaX:2024CEvNS,XENON:2024CEvNS,XENON:2025Fog,LZ:2026CEvNS}. Complementary probes at colliders and through astrophysical observations, spanning electromagnetic, gravitational and cosmic-ray channels, have also become increasingly important in the broader effort to uncover the nature of DM~\cite{Gaskins:2016cha,DeRoeck:2024dm,FermiLAT:2015dSph,HESS:2016Halo,LHAASO:2022HeavyDM,LHAASO:2024UltraheavyDM,Yuan:2022nmu,Leung:2024HAWCDM,Fong:2025eROSITA,Luu:2024WaveDM}.

Inelastic scattering provides a natural mechanism for shifting the recoil spectrum toward higher energies. In this scenario, an incident halo particle $\chi_1$ scatters off a nucleus $N$ into a heavier state through $\chi_1N\to\chi_2N$, with a positive mass splitting $\delta=m_{\chi_2}-m_{\chi_1}$~\cite{TuckerSmith:2001hy,TuckerSmith:2004jv}. The energy required to excite the heavier state raises the minimum incident speed and suppresses low-energy recoils~\cite{XENON100:2011iDM,Bramante:2016rdh,XENON1T:2024EFT}. For splittings of a few hundred keV, the accessible scattering phase space can approach the high-speed endpoint of the Galactic halo. The predicted rate then becomes particularly sensitive to the Earth's motion and to uncertainties in the fastest DM populations, including possible contributions associated with the Large Magellanic Cloud~\cite{Baxter:2021pqo,McCabe:2013kea,Besla:2019lft,SmithOrlik:2023lmc}.

The LZ event has consequently motivated interpretations involving generic two-state DM, Higgsinos, electroweak multiplets, and scalar-doublet models~\cite{Su:2026rwz,Fan:2026kxx,Freese:2026sga,Wu:2026nhi,Yin:2026jnn,Visinelli:2026kgt,Nomura:2026qyq,Smirnov:2026aqk,Du:2026guj}. Several proposed benchmarks lie close to the kinematic boundary~\cite{McCabe:2026crm,Nomura:2026qyq}, while solar capture and the LZ high-energy sideband provide additional tests of Higgsino interpretations~\cite{Pospelov:2026ewn,Rodd:2026tyn}. Other studies have explored momentum- or spin-dependent elastic scattering, inelastic dark-photon DM, fermionic DM absorption, atmospheric-neutrino up-scattering, and related inelastic mechanisms~\cite{DiMauro:2026ldr,Yamashita:2026ump,Lou:2026idn,2026arXiv260904185J,Gu:2026vto,Dent:2026bji}. Together, these proposals illustrate the range of particle-physics and astrophysical assumptions relevant to interpreting a single high-energy recoil.

Axionlike particles (ALPs) and pseudoscalar mediators introduce an additional source of spectral structure through their momentum-dependent scattering amplitudes~\cite{Arina:2014yna,Dolan:2014ska,Abe:2018emu}. Their effective interactions, renormalization-group evolution, and collider and flavor phenomenology have been extensively studied~\cite{Mimasu:2014nea,Bauer:2017ris,Bauer:2020jbp,2022JHEP...09..056B,2024JHEP...01..092B}, and ALPs can mediate interactions between DM and the Standard Model~\cite{Dror:2023fyd}. For the gluonic portal $aG\widetilde G$, Refs~\cite{Bae:2023ago, Beenakker:2025mhf} demonstrated that a multistate Majorana sector permits both elastic and inelastic axion-mediated nuclear scattering. A subsequent treatment of ALP-mediated DM--nucleon scattering incorporated nucleon matching, light-mediator effects, loop-induced spin-independent interactions, and laboratory constraints~\cite{Beenakker:2025mhf}. An elastic axion-portal interpretation of the LZ high-recoil event has also recently been proposed~\cite{Unwin:2026rdp}.

Here we investigate the combined effects of endothermic kinematics and gluonic ALP exchange in a two-state Majorana DM model. The mass splitting suppresses low-energy scattering, while the mediator propagator, nucleon form factors, and nuclear response determine the momentum dependence of the recoil spectrum. Treating the dark-sector and gluonic couplings independently, we identify the masses, splittings, and coupling products compatible with a signal interpretation of LZ230616. The gluonic portal is also subject to flavor, fixed-target, and collider constraints~\cite{Dolan:2014ska,MartinCamalich:2020dfe,Gavela:2019wzg,BaBar:2021ich,NA62:2025hidden,Beenakker:2025mhf}. Our analysis establishes the recoil-based requirements for subsequent tests using the thermal relic abundance and complementary laboratory measurements.

The remainder of this paper is organized as follows. Section~\ref{sec:model} introduces the gluonic ALP portal and the inelastic-scattering calculation. Section~\ref{sec:result} presents the likelihood construction and the results for LZ230616. We discuss their implications and conclude in Sec.~\ref{sec:conclusion}.

\section{Two-state dark matter and Inelastic Scattering}\label{sec:model}
\subsection{Gluonic axion portal}
The dark sector contains two Majorana fermions with $m_{\chi} \equiv m_{\chi_1}$ and $m_{\chi_2}=m_{\chi}+\delta$, where $\delta>0$, and an ALP $a$.
We assume that $\chi_1$ constitutes the dominant dark matter component in the present-day halo. The relevant Lagrangian interactions are
\begin{equation}
\begin{split}
\mathcal L_{\mathrm{int}}={}&-\lambda_\chi(\partial_\mu a)\bar\chi_2\gamma^\mu\gamma^5\chi_1+\mathrm{h.c.}
\\&+\frac{\alpha_s}{8\pi}\lambda_G\,a\,G^A_{\mu\nu}\widetilde G^{A\mu\nu}.
\end{split}
\label{eq:lagrangian}
\end{equation}
Here $G^A_{\mu\nu}$ and $\widetilde G^{A\mu\nu}$ are the gluon field-strength tensor and its dual, respectively, and $\alpha_s$ is the strong coupling. The independent couplings $\lambda_\chi= {c_\chi}/{f_\chi}$ and $\lambda_G={c_G}/{f_G}$ have dimension ${\rm GeV}^{-1}$.
Neither a common decay constant nor the QCD-axion mass coupling relation is imposed, the mediator mass $m_a$ is an independent parameter.

\begin{figure}[t]
\centering
\includegraphics[width=0.7\columnwidth]{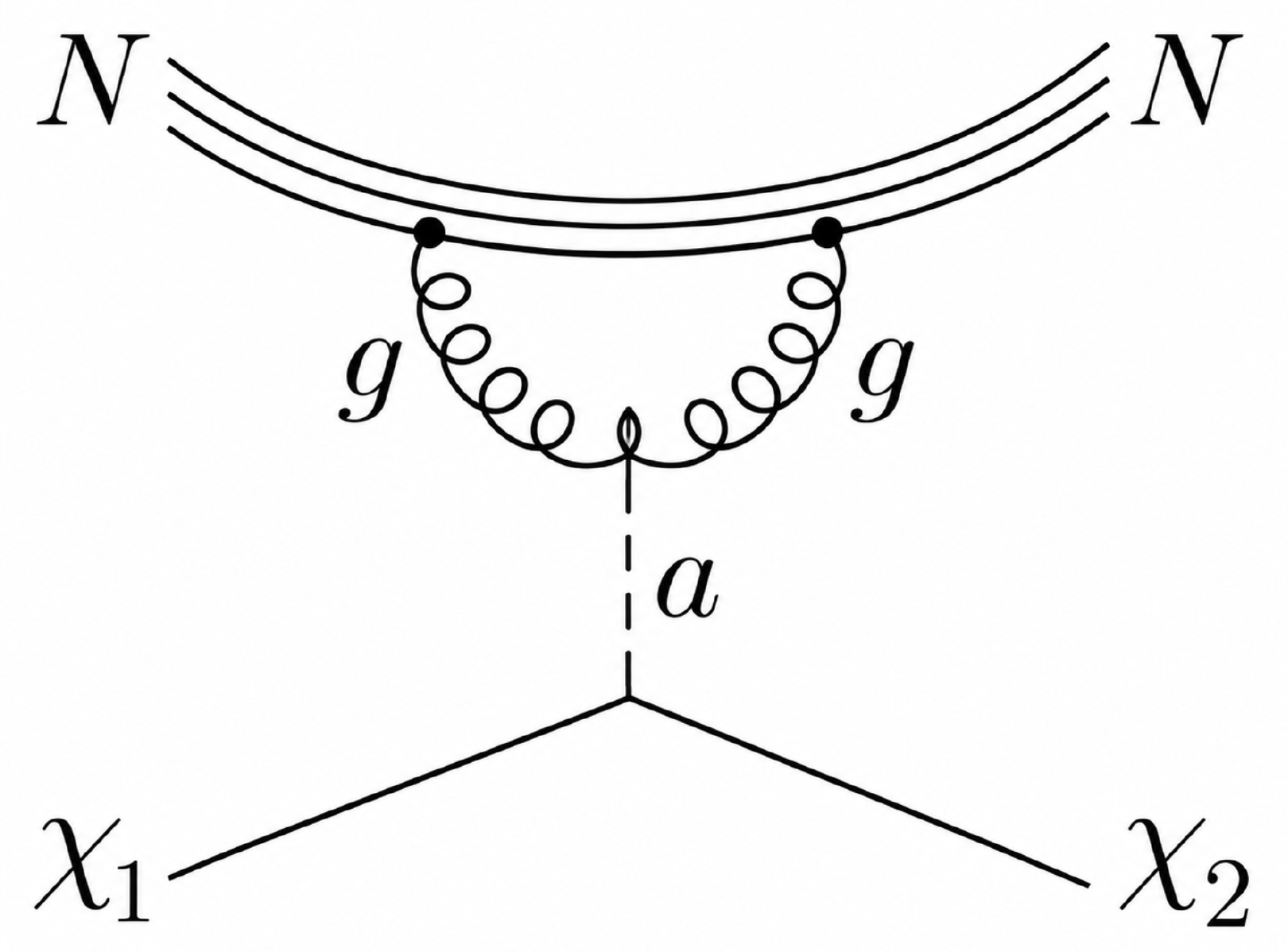}
\caption{Axion-mediated inelastic scattering of two-state dark matter from a nucleon. The incident state is $\chi_1$, and the outgoing state is the slightly heavier $\chi_2$. The ALP couples to the dark sector through an off-diagonal derivative interaction and to the nucleon through the gluonic matrix element of $aG\widetilde G$.}
\label{fig:scattering}
\end{figure}

We consider a minimal two-state Majorana dark sector coupled through a gluonic ALP portal with an off-diagonal interaction. The tree-level elastic process $\chi_1N\rightarrow\chi_1N$ is then absent, whereas the endothermic transition $\chi_1N\rightarrow\chi_2N$ remains allowed, and 
Figure.~\ref{fig:scattering} depicts the resulting transition. At the nucleon vertex, the $aG\widetilde G$ interaction couples the ALP to the gluonic matrix element inside the nucleon. We retain the full nucleon matching and nuclear response, rather than replacing this contribution by a constant nuclear-spin coupling.

The pseudoscalar gluonic matrix element is written as
\begin{equation}
\left\langle N'\left|\frac{\alpha_s}{8\pi}G\widetilde G\right|N\right\rangle
=m_Na_N(q^2)\bar u_N' i\gamma^5u_N.
\end{equation}

We retain the pion and eta pole contributions to $a_N (q^2)$~\cite{Bae:2023ago}, and set the momentum transfer  $q^2=-Q^2<0$.
Keeping the leading pion and eta pole terms, the form factor used in the numerical calculation is therefore
\begin{equation}
a_N(-Q^2)=-\frac{Q^2a_{\widetilde G,\pi}^N}{m_\pi^2+Q^2}
-\frac{Q^2a_{\widetilde G,\eta}^N}{m_\eta^2+Q^2}
+b_{\widetilde G}^N.
\end{equation}
 
The inputs are the proton spin fractions $(\Delta u,\Delta d,\Delta s)_p=(0.847,-0.407,-0.035)$, quark masses $(m_u,m_d,m_s)=(2.16,4.70,93.5)\,\MeV$, and meson masses $m_\pi=134.98\,\MeV$ and $m_\eta=547.86\,\MeV$. Neutron spin fractions follow by interchanging $u$ and $d$.

The dominant nonrelativistic nucleon operator in the numerical calculation is
\begin{equation}
\mathcal O_6=\left(\bm S_\chi\cdot\frac{\bm q}{m_N^{\mathrm{ref}}}\right)
\left(\bm S_N\cdot\frac{\bm q}{m_N^{\mathrm{ref}}}\right).
\end{equation}
where the reference nucleon mass is  $m_N^{\mathrm{ref}}=0.931\,\GeV$ provided in \texttt{WimPyDD}~\cite{Jeong:2021bpl}.
For each nucleon $N=p,n$, the $\mathcal O_6$ Wilson coefficient supplied to the calculation is
\begin{equation}
c_6^N(Q)=4(m_N^{\mathrm{ref}})^2\lprod
\frac{a_N(-Q^2)}{Q^2+m_a^2}, \qquad \lambda_{\chi G} \equiv \lambda_{\chi}\lambda_G.
\label{eq:c6}
\end{equation}

The isospin coefficients obey $c_6^0=c_6^p+c_6^n$ and $c_6^1=c_6^p-c_6^n$; hence $c_6^{p,n}=(c_6^0\pm c_6^1)/2$. The nuclear calculation retains the full one-body density matrices rather than a static spin-expectation approximation. The spin and natural number abundance are $(J,f)=(1/2,0.26401)$ for $^{129}$Xe and $(3/2,0.21232)$ for $^{131}$Xe. Their numbers per kilogram of natural xenon are $n_{129}=1.21\times10^{24}\,\mathrm{kg}^{-1}$ and $n_{131}=9.73\times10^{23}\,\mathrm{kg}^{-1}$.
These numbers already include the natural abundances. The cross section is evaluated separately for each isotope before applying its weight; no additional abundance factor multiplies $n_T$.

\subsection{Inelastic Scattering and Event Rate in LZ} \label{sec:inelastic_kinematics}

For scattering from a target nucleus $N$, the minimum incident speed required to produce a recoil energy $E_R$ is
\begin{equation}
v_{\min}(E_R)=\frac{m_N E_R/\mu_{\chi N}+\delta}{\sqrt{2m_NE_R}}, \qquad \mu_{\chi N}=\frac{m_\chi m_N}{m_\chi+m_N}.
\end{equation}
Its minimum occurs at
\begin{equation}
E_R^*=\frac{\mu_{\chi N}}{m_N}\delta,\qquad
v_{\min}^*=\sqrt{\frac{2\delta}{\mu_{\chi N}}}.
\end{equation}

Requiring this kinematically favored recoil energy to lie near the LZ candidate energy, $E_R\simeq248\,\keV$, gives $\delta\simeq248\,\keV\,({m_N}/{\mu_{\chi N}})$.
Thus DM masses of several hundred GeV to a TeV point to splittings of a few hundred keV. This relation is not a strict mass bound: recoils away from $E_R^*$ remain possible, at the cost of a larger required speed or coupling.

We calculate the rate using the Standard Halo Model with $v_0=238\,\mathrm{km\,s}^{-1},
v_{\mathrm{esc}}=544\,\mathrm{km\,s}^{-1}$, and $\rho_\chi=0.3\,\GeV\,\mathrm{cm}^{-3}$, averaging the inverse-speed integral over a full year.
For the normalized Earth-frame velocity distribution $f_\oplus(\bm v,t)$, we define
For the normalized Earth-frame velocity distribution $f_\oplus(\bm v,t)$, we define
\begin{align}
\eta(v_{\min},t)&=\int_{|\bm v|\ge v_{\min}}\frac{f_\oplus(\bm v,t)}{|\bm v|}\,\dd^3v,\\
\bar\eta(v_{\min})&=\frac{1}{T_{\mathrm{yr}}}\int_0^{T_{\mathrm{yr}}}\eta(v_{\min},t)\,\dd t.
\end{align}
The Galactic distribution is the truncated Maxwellian of the Standard Halo Model, with the parameters quoted in the main text. The full-year average is an approximation to the exposure-weighted halo integral. 
The Earth-frame speed boundary is $v_{\mathrm{esc}}+|\bm v_\oplus(t)|$, so $v_{\mathrm{esc}}$ alone cannot be used as the maximum incident laboratory speed.

\begin{figure}[htbp]
\centering
\includegraphics[width=0.98\linewidth]{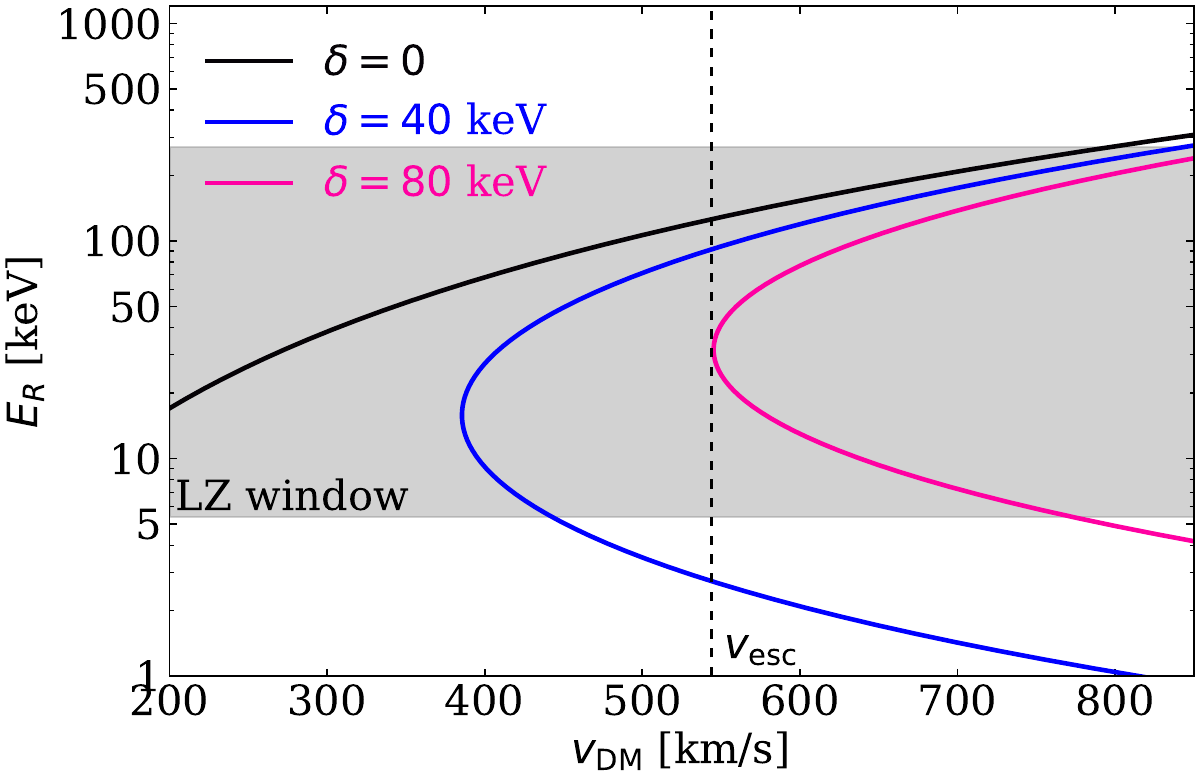}
\caption{Minimum speed and the high-recoil energy scale. The available range of recoil energies on a nuclear target with a DM mass at $m_{\chi} = 100$~GeV. The gray-shaded region indicates the recoil-energy window measured by LZ. 
The vertical dotted line marks the Galactic-frame escape speed, $v_{\rm esc}$. The maximum DM speed in the laboratory frame can exceed $v_{\rm esc}$ due to the Earth's motion relative to the Galactic halo. }
\label{fig:recoil_energy}
\end{figure}

For a natural-xenon target, the differential event rate is
\begin{equation}
\frac{\dd R}{\dd E_R}=\frac{\rho_\chi}{m_\chi}
\sum_{N=129,131}n_T\int\dd^3v\,f_\oplus(\bm v,t)\,v
\frac{\dd\sigma}{\dd E_R},
\end{equation}
where $\dd\sigma/\dd E_R$ is calculated using Eq.~\eqref{eq:c6} and the full \texttt{WimPyDD} nuclear response. The rate is averaged over a year as described above.

The efficiency-weighted expected recoil spectrum is
\begin{equation}
\frac{\dd N}{\dd E_R}=\mathcal E\,\epsilon_{\mathrm{LZ}}(E_R)
\frac{\dd R}{\dd E_R}.
\end{equation}

The expected spectrum includes the published LZ energy-dependent efficiency and the $\mathcal E=2.84\,\mathrm{tonne\,yr}$ exposure. 


\section{Likelihood and Model Fitting}\label{sec:result}
\subsection{Likelihood normalization}
To separate spectral compatibility from interaction strength, let $s(E_R|\theta)$ be the expected spectrum at a reference coupling, with $\theta=(m_\chi,\delta,m_a)$, and let $A$ scale its normalization. Define $S=\int_{\mathcal W}s\,\dd E_R$ and $J=\int_{\mathcal W}sG(E_{\mathrm{obs}}|E_R,\sigma_E)\,\dd E_R$, where $G$ is a normalized Gaussian and $\sigma_E=\sqrt{23^2+23^2}\,\keV=32.53\,\keV$. Under the conditional signal-only, single-event hypothesis,
\begin{equation}
\mathcal L(A,\theta)\propto e^{-AS}AJ,\qquad
\widehat A=\frac1S,\qquad\mathcal L_{\mathrm{prof}}\propto\frac JS.
\label{eq:likelihood}
\end{equation}
The relative statistic is
\begin{equation}
\qLZ=-2\ln\frac{\mathcal L_{\mathrm{prof}}(\theta)}{\mathcal L_{\mathrm{prof,max}}}.
\label{eq:qlz}
\end{equation}
This recoil-energy proxy does not reproduce LZ's likelihood in detector observables, and its thresholds are not coverage-calibrated confidence regions. It assesses spectral compatibility conditional on a signal interpretation, not the probability that the event is DM. Since the rate scales as $|\lprod|^2$, the coupling giving one expected event follows from $|\lprod|=|\lambda_{\chi G}^{\mathrm{ref}}|/\sqrt{N_{\mathrm{ref}}}$.

\begin{figure*}[htbp]
\centering
\includegraphics[width=0.75\textwidth]{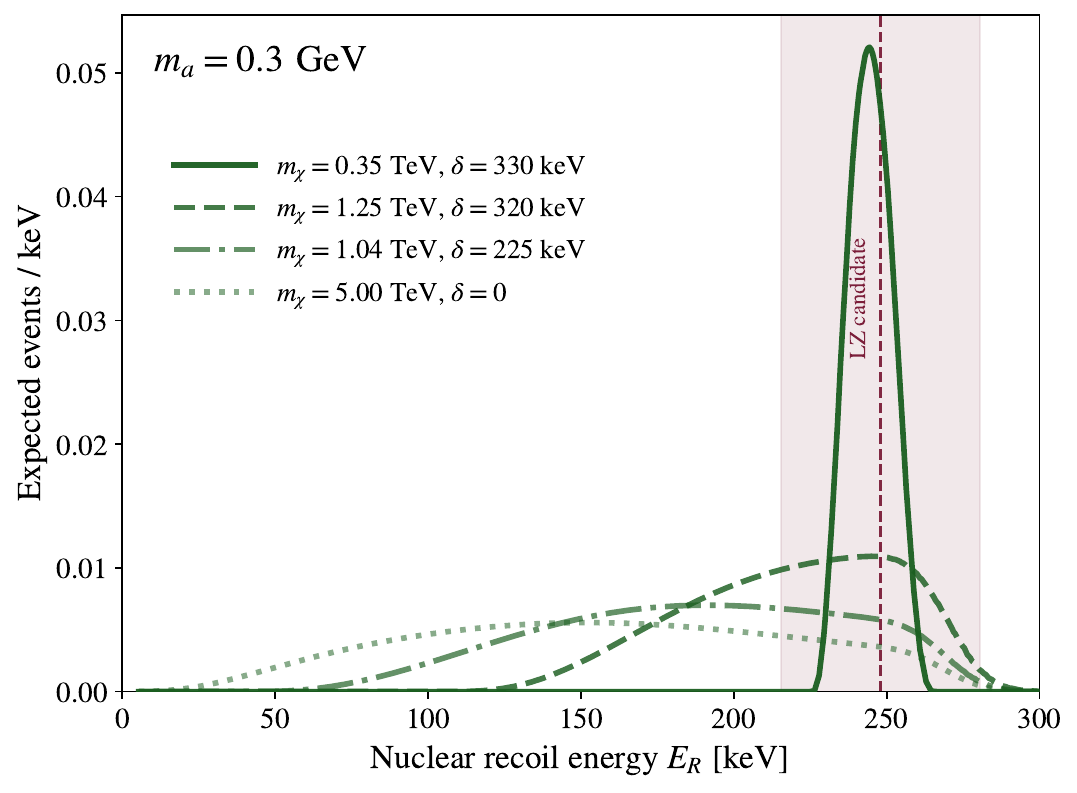}
\caption{Expected recoil spectra for $m_a=0.3\,\GeV$, each normalized to one event in $5.4$--$269.9\,\keV$. The solid curve gives the best spectral shape, the long-dashed and dash-dotted curves minimize the required coupling among scanned points with $\qLZ\le1$ and $\qLZ\le2$, respectively. The dotted curve is the degenerate-mass limit of the same off-diagonal model. The vertical line marks $248\,\keV$; the band shows the Gaussian energy proxy, not an official LZ likelihood interval. The extension to $300\,\keV$ displays only the spectral tails.}
\label{fig:spectra}
\end{figure*}

At $\lambda_{\chi G}^{\mathrm{ref}}=10^{-6}\,\GeV^{-2}$, let $s(E_R|\theta)$ include the efficiency and exposure. It is an expected count density, not a unit-area probability distribution. The normalization parameter is
\begin{equation}
A=\left|\frac{\lprod}{\lambda_{\chi G}^{\mathrm{ref}}}\right|^2\ge0,
\end{equation}
and the total expected signal is $AS$. The energy kernel is
\begin{equation}
G(E_{\mathrm{obs}}|E_R,\sigma_E)=\frac{1}{\sqrt{2\pi}\sigma_E}
\exp\left[-\frac{(E_{\mathrm{obs}}-E_R)^2}{2\sigma_E^2}\right].
\end{equation}
It approximates the reported statistical and systematic energy uncertainties as independent Gaussian contributions. This kernel is not a reconstruction of LZ's detector response or official event-energy likelihood.

For one observed event and zero modeled background, the Poisson factor $e^{-AS}AS$ times the event-energy density $J/S$ gives Eq.~\eqref{eq:likelihood}. For $S>0$, maximization over $A$ yields $\widehat A=1/S$, so every profiled spectrum predicts one event in $\mathcal W$. The residual likelihood compares the energy distributions through $J/S$. The maximum in Eq.~\eqref{eq:qlz} is taken over the scanned parameters; thresholds $\qLZ\le1$ and $\qLZ\le2$ label relative spectral agreement only. No background significance or frequentist coverage is inferred from them.

The coupling products quoted in the main text use this conditional normalization. They do not include a relic-abundance prior or collider and flavor constraints. In particular, the $350.2\,\GeV$, $330\,\keV$ point depends on a very small accessible part of the high-speed halo and should be interpreted together with its endpoint sensitivity. The broader TeV-scale spectra provide alternatives with much smaller normalization requirements, without resolving the remaining particle-physics consistency questions.

\subsection{Model Fitting with LZ230616}\label{sec:implication}

For the scanned mediator masses $m_a=0.1$, $0.3$, $1$, and $3\,\GeV$, the best spectral-shape point lies near $m_\chi=350\,\GeV$ and $\delta=330\,\keV$. Its stability across these mediator masses indicates that the inelastic threshold mainly sets the peak position; the propagator primarily changes the required normalization. Figure~\ref{fig:spectra} illustrates the result for $m_a=0.3\,\GeV$, with each spectrum normalized to one expected event in $\mathcal W$.

The solid curve is narrowly concentrated near the candidate energy because only particles close to the maximum halo speed can scatter. At $(m_\chi,\delta)=(350.2\,\GeV,330\,\keV)$, producing one expected event requires $|\lprod|\simeq1.51\,\GeV^{-2}$. The minimum speeds at $E_R=248\,\keV$ are approximately $814.5$ and $809.9\,\mathrm{km\,s}^{-1}$ for $^{129}$Xe and $^{131}$Xe, respectively. In the adopted annual treatment the point is supported predominantly by the summer phase space of $^{131}$Xe, while the $^{129}$Xe channel is closed. Small changes in the speed endpoint can then cause large changes in the inferred coupling. This is an extreme spectral benchmark, rather than an established viable particle-physics solution.

\begin{table}[htbp]
\centering
\caption{Benchmark parameters and spectral-compatibility statistics
for an axionlike mediator mass $m_a=0.3\,\mathrm{GeV}$ displayed in Figure~\ref{fig:spectra}.}
\label{tab:benchmarks-ma03}
\begin{tabular}{lrrrr}
\hline\hline
Benchmark
& $m_\chi\;[\mathrm{TeV}]$
& $\delta\;[\mathrm{keV}]$
& $\Delta q_{\mathrm{LZ}}$
& $|\lambda_\chi\lambda_G|\;[\mathrm{GeV}^{-2}]$ \\
\hline
(a) Best fit
& $0.35$ & $330$ & $0$
& $0.21$ \\

(b) $\Delta q_{\mathrm{LZ}}\leq 1$
& $1.25$ & $320$ & $0.99$
& $2.30\times10^{-4}$ \\

(c) $\Delta q_{\mathrm{LZ}}\leq 2$
& $1.04$ & $225$ & $1.99$
& $4.30\times10^{-5}$ \\

(d) Elastic axion
& $5.0$ & $0$ & $2.84$
& $1.42\times10^{-5}$ \\
\hline\hline
\end{tabular}
\end{table}

Allowing a modest degradation in spectral agreement changes the normalization dramatically. 
For \(m_a=0.3\) GeV, the best spectral-shape point is located at \((m_\chi,\delta)\simeq(350.2~{\rm GeV},330~{\rm keV})\), requiring \(|\lambda_{\chi G}|\simeq 0.214~{\rm GeV}^{-2}\) for one expected event in the ROI. Subject to \(\Delta q_{\rm LZ}\le1\), the minimum-coupling point shifts to \((m_\chi,\delta)\simeq(1.25~{\rm TeV},320~{\rm keV})\), with \(|\lambda_{\chi G}|\simeq2.30\times10^{-4}~{\rm GeV}^{-2}\). Relaxing the spectral criterion to \(\Delta q_{\rm LZ}\le2\) gives \((m_\chi,\delta)\simeq(1.04~{\rm TeV},225~{\rm keV})\), requiring \(|\lambda_{\chi G}|\simeq4.30\times10^{-5}~{\rm GeV}^{-2}\).
Both xenon channels have open phase space throughout the year at these alternatives. The spectra are broader, but the required couplings are smaller by more than three orders of magnitude and less sensitive to the endpoint. By comparison, the $\delta=0$ reference retains a broad distribution at lower energies and does not generate the same sharp localization.

The event count alone therefore does not isolate a narrow region in $(m_\chi,\delta)$. At fixed normalization it would constrain both parameters, as illustrated by the fixed-cross-section study of Ref.~\cite{Su:2026rwz}. With an independent coupling product, profiling instead assigns the count to $|\lprod|$ and leaves a degeneracy between mass and splitting. Likewise, the thermal mass selection in Ref.~\cite{Wu:2026nhi} supplies information absent from an LZ-only analysis of the gluonic portal. These distinctions are essential when comparing apparently similar TeV-scale benchmarks.

\section{Discussion and Conclusions}\label{sec:conclusion}

We have investigated axion-mediated inelastic scattering between two Majorana dark matter states as a possible interpretation of LZ230616.
A positive mass splitting suppresses low-energy recoils, while the momentum-dependent gluonic form factor and xenon nuclear response shape the high-energy spectrum. Conditional on a signal interpretation, the recoil spectrum constrains the relation between $m_\chi$ and $\delta$, and the event normalization determines $|\lambda_\chi\lambda_G|$.

For $m_a=0.3\,\mathrm{GeV}$, the best spectral-shape point near $m_\chi=350\,\mathrm{GeV}$ and $\delta=330\,\mathrm{keV}$ produces a narrow peak close to $248\,\mathrm{keV}$, but requires a large coupling product and is highly sensitive to the halo speed endpoint.
TeV-scale benchmarks with moderately smaller splittings reduce the required coupling product by several orders of magnitude and alleviate endpoint sensitivity with only a modest loss of spectral agreement. 
We emphasize that our statistic, $\Delta q_{\mathrm{LZ}}$, is constructed from an approximate single-event recoil-energy likelihood and therefore neither reproduces the official LZ likelihood nor defines statistically calibrated confidence regions. Moreover, we treat the present-day dominance of $\chi_1$ as an assumption of the recoil analysis, leaving a detailed assessment of its consistency with the cosmological evolution of the two-state system for future work.

A single candidate therefore cannot establish a unique or fully viable dark matter solution. A self-consistent interpretation requires a two-state relic-abundance calculation to constrain $\lambda_\chi$, followed by collider, rare-meson-decay, and other laboratory tests of the inferred $\lambda_G$. Any surviving parameter space would provide predictions for future high-energy recoil searches and indirect detection, including effects associated with the abundance and dynamics of the heavier state.

\vspace{0.8em}
\textbf{\textit{Acknowledgements.}} This work is supported by the National Natural Science Foundation of China under grants 12373002, 12393854 and 12588101.


\vspace{0.8em}
{\footnotesize
\noindent$^{*}$ guanwen.yuan@ustc.edu.cn\\
$^{\dagger}$ zhangbo@pmo.ac.cn\\
$^{\ddagger}$ wenyucao@cuhk.edu.hk\\
$^{\S}$ fenglei@pmo.ac.cn\\
$^{\P}$ yangrz@ustc.edu.cn
}

\bibliography{reference}

@article{LZ:2026axp,
    author = "Akerib, D. S. and others",
    collaboration = "LZ",
    title = "{Search for dark matter particle interactions in an extended nuclear recoil energy window with the LUX-ZEPLIN (LZ) experiment}",
    eprint = "2609.02823",
    archivePrefix = "arXiv",
    primaryClass = "hep-ex",
    month = "9",
    year = "2026"
}

@article{Anand:2013yka,
    author = "Anand, Nikhil and Fitzpatrick, A. Liam and Haxton, W. C.",
    title = "{Weakly interacting massive particle-nucleus elastic scattering response}",
    eprint = "1308.6288",
    archivePrefix = "arXiv",
    primaryClass = "hep-ph",
    doi = "10.1103/PhysRevC.89.065501",
    journal = "Phys. Rev. C",
    volume = "89",
    number = "6",
    pages = "065501",
    year = "2014"
}

@article{Yuan:2022nmu,
    author = "Yuan, Guan-Wen and Shen, Zhao-Qiang and Tsai, Yue-Lin Sming and Yuan, Qiang and Fan, Yi-Zhong",
    title = "{Constraining ultralight bosonic dark matter with Keck observations of S2{\textquoteright}s orbit and kinematics}",
    eprint = "2205.04970",
    archivePrefix = "arXiv",
    primaryClass = "astro-ph.HE",
    doi = "10.1103/PhysRevD.106.103024",
    journal = "Phys. Rev. D",
    volume = "106",
    number = "10",
    pages = "103024",
    year = "2022"
}

@article{Lou:2026idn,
    author = "Lou, Yuanchao and Lu, Chih-Ting",
    title = "{Fermionic Dark Matter Absorption and the High-Energy Event in LUX-ZEPLIN}",
    eprint = "2609.01592",
    archivePrefix = "arXiv",
    primaryClass = "hep-ph",
    month = "9",
    year = "2026"
}

@article{Su:2026rwz,
    author = "Su, Liangliang and Yang, Jin Min and Yang, Wen-Na",
    title = "{Inelastic Dark Matter Signature at High Recoil Energy in LUX-ZEPLIN and CRESST}",
    eprint = "2609.01475",
    archivePrefix = "arXiv",
    primaryClass = "hep-ph",
    month = "9",
    year = "2026"
}

@article{Yamashita:2026ump,
    author = "Yamashita, Kimiko",
    title = "{Inelastic Dark Photon Dark Matter for the LUX-ZEPLIN High-Recoil Event and the Galactic Halo Gamma-Ray Excess}",
    eprint = "2609.02868",
    archivePrefix = "arXiv",
    primaryClass = "hep-ph",
    month = "9",
    year = "2026"
}

@article{Fan:2026kxx,
    author = "Fan, JiJi and Reece, Matthew",
    title = "{Higgsino Above the Sea of Fog}",
    eprint = "2609.01504",
    archivePrefix = "arXiv",
    primaryClass = "hep-ph",
    month = "9",
    year = "2026"
}

@article{Freese:2026sga,
    author = "Freese, Katherine and Theodosopoulos, Dionysios P.",
    title = "{Higgsino Dark Matter Interpretation of the LUX-ZEPLIN 248 keV Nuclear-Recoil Event}",
    eprint = "2609.01583",
    archivePrefix = "arXiv",
    primaryClass = "hep-ph",
    month = "9",
    year = "2026"
}

@article{Wu:2026nhi,
    author = "Wu, Lei and Zhang, Yang and Zhu, Bin",
    title = "{TeV Higgsino Dark Matter from LZ Nuclear Recoil to Fermi-LAT Gamma Rays}",
    eprint = "2609.01590",
    archivePrefix = "arXiv",
    primaryClass = "hep-ph",
    month = "9",
    year = "2026"
}

@article{Yin:2026jnn,
    author = "Yin, Wen",
    title = "{A PQ-Symmetric High-Scale SUSY Interpretation of the LZ High-Energy Recoil}",
    eprint = "2609.01892",
    archivePrefix = "arXiv",
    primaryClass = "hep-ph",
    month = "9",
    year = "2026"
}

@article{DiMauro:2026ldr,
    author = "Di Mauro, Mattia",
    title = "{Dark Matter at the Kinematic Edge: Interpreting the 248 keV LZ Nuclear-Recoil Candidate}",
    eprint = "2609.02608",
    archivePrefix = "arXiv",
    primaryClass = "hep-ph",
    month = "9",
    year = "2026"
}

@article{Visinelli:2026kgt,
    author = "Visinelli, Luca",
    title = "{A Peccei--Quinn Origin for Inelastic Electroweak Dark Matter after LUX-ZEPLIN}",
    eprint = "2609.02807",
    archivePrefix = "arXiv",
    primaryClass = "hep-ph",
    month = "9",
    year = "2026"
}

@article{Unwin:2026rdp,
    author = "Unwin, James",
    title = "{Axion Portal Dark Matter and the LUX-ZEPLIN High-Recoil Event}",
    eprint = "2609.04186",
    archivePrefix = "arXiv",
    primaryClass = "hep-ph",
    month = "9",
    year = "2026"
}

@article{Beenakker:2025mhf,
    author = {Beenakker, Wim and Mikkers, Dani{\"e}l and Phan, Anh Vu and Westhoff, Susanne},
    title = "{ALP-mediated dark matter-nucleon scattering}",
    eprint = "2511.19619",
    archivePrefix = "arXiv",
    primaryClass = "hep-ph",
    reportNumber = "Nikhef 2025-017",
    doi = "10.1007/JHEP05(2026)033",
    journal = "JHEP",
    volume = "05",
    pages = "033",
    year = "2026"
}

@article{Bae:2023ago,
    author = "Bae, Kyu Jung and Kim, Jongkuk",
    title = "{Axion-Mediated Inelastic Dark Matter}",
    eprint = "2312.11210",
    archivePrefix = "arXiv",
    primaryClass = "hep-ph",
    month = "12",
    year = "2023"
}

@ARTICLE{TuckerSmith:2001hy,
       author = {{Tucker-Smith}, David and {Weiner}, Neal},
        title = "{Inelastic dark matter}",
      journal = {\prd},
         year = 2001,
        month = jul,
       volume = {64},
       number = {4},
          eid = {043502},
        pages = {043502},
          doi = {10.1103/PhysRevD.64.043502},
archivePrefix = {arXiv},
       eprint = {hep-ph/0101138},
 primaryClass = {hep-ph},
       adsurl = {https://ui.adsabs.harvard.edu/abs/2001PhRvD..64d3502T}
}

@ARTICLE{Bramante:2016rdh,
       author = {{Bramante}, Joseph and {Fox}, Patrick J. and {Kribs}, Graham D. and {Martin}, Adam},
        title = "{The Inelastic Frontier: Discovering Dark Matter at High Recoil Energy}",
      journal = {\prd},
         year = 2016,
        month = dec,
       volume = {94},
       number = {11},
          eid = {115026},
        pages = {115026},
          doi = {10.1103/PhysRevD.94.115026},
archivePrefix = {arXiv},
       eprint = {1608.02662},
 primaryClass = {hep-ph},
       adsurl = {https://ui.adsabs.harvard.edu/abs/2016PhRvD..94k5026B}
}

@ARTICLE{2026arXiv260904185J,
       author = {{Jeesun}, Sk and {Majumdar}, Anirban},
        title = "{Atmospheric neutrino up-scattering explanation of LZ 2026 excess}",
      journal = {arXiv e-prints},
         year = 2026,
        month = sep,
          eid = {arXiv:2609.04185},
        pages = {arXiv:2609.04185},
archivePrefix = {arXiv},
       eprint = {2609.04185},
 primaryClass = {hep-ph},
       adsurl = {https://ui.adsabs.harvard.edu/abs/2026arXiv260904185J}
}

@article{Dent:2026bji,
    author = "Dent, James B. and Newstead, Jayden L.",
    title = "{Exothermic and Endothermic Inelastic Dark Matter Interpretations at LZ: Sideband Constraints and Future Prospects}",
    eprint = "2609.04673",
    archivePrefix = "arXiv",
    primaryClass = "hep-ph",
    month = "9",
    year = "2026"
}

@article{Gu:2026vto,
    author = "Gu, Guanhua and Li, Lingfeng and Tang, Shao-Song and Xu, Yongheng",
    title = "{Inelastic from the Other Side: Xenon Excitation Signals in Light of the LZ High-Recoil Event}",
    eprint = "2609.05291",
    archivePrefix = "arXiv",
    primaryClass = "hep-ph",
    month = "9",
    year = "2026"
}

@article{Smirnov:2026aqk,
    author = "Smirnov, Juri and Griffith, Spencer and Beacom, John F.",
    title = "{Inelastic Signatures of Electroweak Dark Matter}",
    eprint = "2609.04144",
    archivePrefix = "arXiv",
    primaryClass = "hep-ph",
    month = "9",
    year = "2026"
}

@article{Du:2026guj,
    author = "Du, Xiaokang and Wang, Fei",
    title = "{TeV Higgsino Interpretation of the LZ High-Recoil Event with Intermediate-Scale Electroweak Gauginos}",
    eprint = "2609.04163",
    archivePrefix = "arXiv",
    primaryClass = "hep-ph",
    month = "9",
    year = "2026"
}

@article{McCabe:2026crm,
    author = "McCabe, Christopher",
    title = "{Seasonal dark matter from the LUX-ZEPLIN high-energy event}",
    eprint = "2609.04181",
    archivePrefix = "arXiv",
    primaryClass = "hep-ph",
    month = "9",
    year = "2026"
}

@article{Rodd:2026tyn,
    author = "Rodd, Nicholas L. and Safdi, Benjamin R. and Slatyer, Tracy R. and Xu, Weishuang Linda",
    title = "{Confronting the Higgsino Interpretation of the LZ Event with the High-Energy Sideband}",
    eprint = "2609.04175",
    archivePrefix = "arXiv",
    primaryClass = "hep-ph",
    month = "9",
    year = "2026"
}

@article{Pospelov:2026ewn,
    author = "Pospelov, Maxim and Ramani, Harikrishnan",
    title = "{Strong Constraints on Higgsino Dark Matter from Solar Capture}",
    eprint = "2609.02775",
    archivePrefix = "arXiv",
    primaryClass = "hep-ph",
    month = "9",
    year = "2026"
}

@article{Nomura:2026qyq,
    author = "Nomura, Yasunori",
    title = "{Dark Matter as the Z{\_}2 Partner of the Standard Model Higgs Boson}",
    eprint = "2609.02505",
    archivePrefix = "arXiv",
    primaryClass = "hep-ph",
    reportNumber = "RIKEN-iTHEMS-Report-26",
    month = "9",
    year = "2026"
}

@article{Jeong:2021bpl,
    author = "Jeong, Injun and Kang, Sunghyun and Scopel, Stefano and Tomar, Gaurav",
    title = "{WimPyDD: An object{\textendash}oriented Python code for the calculation of WIMP direct detection signals}",
    eprint = "2106.06207",
    archivePrefix = "arXiv",
    primaryClass = "hep-ph",
    reportNumber = "CQUeST-2021-0663, TUM-HEP 1343/21",
    doi = "10.1016/j.cpc.2022.108342",
    journal = "Comput. Phys. Commun.",
    volume = "276",
    pages = "108342",
    year = "2022"
}

@ARTICLE{2022JHEP...09..056B,
       author = {{Bauer}, Martin and {Neubert}, Matthias and {Renner}, Sophie and {Schnubel}, Marvin and {Thamm}, Andrea},
        title = "{Flavor probes of axion-like particles}",
      journal = {Journal of High Energy Physics},
         year = 2022,
        month = sep,
       volume = {2022},
       number = {9},
          eid = {056},
        pages = {056},
          doi = {10.1007/JHEP09(2022)056},
archivePrefix = {arXiv},
       eprint = {2110.10698},
 primaryClass = {hep-ph},
       adsurl = {https://ui.adsabs.harvard.edu/abs/2022JHEP...09..056B}
}

@ARTICLE{2024JHEP...01..092B,
       author = {{Bruggisser}, Sebastian and {Grabitz}, Lara and {Westhoff}, Susanne},
        title = "{Global analysis of the ALP effective theory}",
      journal = {Journal of High Energy Physics},
         year = 2024,
        month = jan,
       volume = {2024},
       number = {1},
          eid = {092},
        pages = {092},
          doi = {10.1007/JHEP01(2024)092},
archivePrefix = {arXiv},
       eprint = {2308.11703},
 primaryClass = {hep-ph},
       adsurl = {https://ui.adsabs.harvard.edu/abs/2024JHEP...01..092B}
}

@ARTICLE{Balazs:2024Primer,
       author = {{Balazs}, Csaba and {Bringmann}, Torsten and {Kahlhoefer}, Felix and {White}, Martin},
        title = "{A Primer on Dark Matter}",
      journal = {Astrophysics},
         year = 2026,
       volume = {5},
        pages = {17},
          doi = {10.1016/B978-0-443-21439-4.00070-5},
archivePrefix = {arXiv},
       eprint = {2411.05062},
 primaryClass = {astro-ph.CO}
}

@ARTICLE{Xia:2026DirectDM,
       author = {{Xia}, Qing and {Canonica}, Lucia},
        title = "{Progress and prospects in the underground laboratories' search for dark matter}",
      journal = {Communications Physics},
         year = 2026,
       volume = {9},
          eid = {105},
        pages = {105},
          doi = {10.1038/s42005-026-02563-1}
}

@ARTICLE{LZ:2025WIMP,
       author = {{Aalbers}, J. and others},
 collaboration = {LZ Collaboration},
        title = "{Dark Matter Search Results from 4.2 Tonne-Years of Exposure of the LUX-ZEPLIN (LZ) Experiment}",
      journal = {\prl},
         year = 2025,
       volume = {135},
          eid = {011802},
        pages = {011802},
          doi = {10.1103/4dyc-z8zf},
archivePrefix = {arXiv},
       eprint = {2410.17036},
 primaryClass = {hep-ex}
}

@ARTICLE{PandaX:2025WIMP,
       author = {{Bo}, Zihao and others},
 collaboration = {PandaX Collaboration},
        title = "{Dark Matter Search Results from 1.54 Tonne$\cdot$Year Exposure of PandaX-4T}",
      journal = {\prl},
         year = 2025,
       volume = {134},
          eid = {011805},
        pages = {011805},
          doi = {10.1103/PhysRevLett.134.011805},
archivePrefix = {arXiv},
       eprint = {2408.00664},
 primaryClass = {hep-ex}
}

@ARTICLE{XENON:2025WIMP,
       author = {{Aprile}, E. and others},
 collaboration = {XENON Collaboration},
        title = "{WIMP Dark Matter Search Using a 3.1 Tonne-Year Exposure of the XENONnT Experiment}",
      journal = {\prl},
         year = 2025,
       volume = {135},
          eid = {221003},
        pages = {221003},
          doi = {10.1103/msw4-t342},
archivePrefix = {arXiv},
       eprint = {2502.18005},
 primaryClass = {hep-ex}
}

@ARTICLE{PandaX:2024CEvNS,
       author = {{Bo}, Zihao and others},
 collaboration = {PandaX Collaboration},
        title = "{First Indication of Solar ${}^{8}\mathrm{B}$ Neutrinos through Coherent Elastic Neutrino-Nucleus Scattering in PandaX-4T}",
      journal = {\prl},
         year = 2024,
       volume = {133},
          eid = {191001},
        pages = {191001},
          doi = {10.1103/PhysRevLett.133.191001},
archivePrefix = {arXiv},
       eprint = {2407.10892},
 primaryClass = {hep-ex}
}

@ARTICLE{XENON:2024CEvNS,
       author = {{Aprile}, E. and others},
 collaboration = {XENON Collaboration},
        title = "{First Indication of Solar ${}^{8}\mathrm{B}$ Neutrinos via Coherent Elastic Neutrino-Nucleus Scattering with XENONnT}",
      journal = {\prl},
         year = 2024,
       volume = {133},
          eid = {191002},
        pages = {191002},
          doi = {10.1103/PhysRevLett.133.191002},
archivePrefix = {arXiv},
       eprint = {2408.02877},
 primaryClass = {hep-ex}
}

@ARTICLE{XENON:2025Fog,
       author = {{Aprile}, E. and others},
 collaboration = {XENON Collaboration},
        title = "{First Search for Light Dark Matter in the Neutrino Fog with XENONnT}",
      journal = {\prl},
         year = 2025,
       volume = {134},
          eid = {111802},
        pages = {111802},
          doi = {10.1103/PhysRevLett.134.111802},
archivePrefix = {arXiv},
       eprint = {2409.17868},
 primaryClass = {hep-ex}
}

@ARTICLE{LZ:2026CEvNS,
       author = {{Akerib}, D. S. and others},
 collaboration = {LZ Collaboration},
        title = "{Searches for Light Dark Matter and Evidence of Coherent Elastic Neutrino-Nucleus Scattering of Solar Neutrinos with the LUX-ZEPLIN (LZ) Experiment}",
      journal = {\prl},
         year = 2026,
       volume = {137},
          eid = {091806},
        pages = {091806},
          doi = {10.1103/jvqf-njpj},
archivePrefix = {arXiv},
       eprint = {2512.08065},
 primaryClass = {hep-ex}
}

@ARTICLE{LHAASO:2022HeavyDM,
       author = {{Cao}, Zhen and others},
 collaboration = {LHAASO Collaboration},
        title = "{Constraints on Heavy Decaying Dark Matter from 570 Days of LHAASO Observations}",
      journal = {\prl},
         year = 2022,
       volume = {129},
          eid = {261103},
        pages = {261103},
          doi = {10.1103/PhysRevLett.129.261103},
archivePrefix = {arXiv},
       eprint = {2210.15989},
 primaryClass = {astro-ph.HE}
}

@ARTICLE{LHAASO:2024UltraheavyDM,
       author = {{Cao}, Zhen and others},
 collaboration = {LHAASO Collaboration},
        title = "{Constraints on Ultraheavy Dark Matter Properties from Dwarf Spheroidal Galaxies with LHAASO Observations}",
      journal = {\prl},
         year = 2024,
       volume = {133},
          eid = {061001},
        pages = {061001},
          doi = {10.1103/PhysRevLett.133.061001},
archivePrefix = {arXiv},
       eprint = {2406.08698},
 primaryClass = {astro-ph.HE}
}

@ARTICLE{Leung:2024HAWCDM,
       author = {{Leung}, Dylan M. H. and {Ng}, Kenny C. Y.},
        title = "{Improving HAWC dark matter constraints with inverse-Compton emission}",
      journal = {\prd},
         year = 2024,
       volume = {110},
          eid = {103021},
        pages = {103021},
          doi = {10.1103/PhysRevD.110.103021},
archivePrefix = {arXiv},
       eprint = {2312.08989},
 primaryClass = {astro-ph.HE}
}

@ARTICLE{Fong:2025eROSITA,
       author = {{Fong}, Chingam and {Ng}, Kenny C. Y. and {Liu}, Qishan},
        title = "{Searching for particle dark matter with eROSITA early data}",
      journal = {\prd},
         year = 2025,
       volume = {112},
          eid = {083052},
        pages = {083052},
          doi = {10.1103/f2dq-hq2j},
archivePrefix = {arXiv},
       eprint = {2401.16747},
 primaryClass = {astro-ph.HE}
}

@ARTICLE{Luu:2024WaveDM,
       author = {{Luu}, Hoang Nhan and {Liu}, Tao and {Ren}, Jing and {Broadhurst}, Tom and {Yang}, Ruizhi and {Wang}, Jie-Shuang and {Xie}, Zhen},
        title = "{Stochastic Wave Dark Matter with Fermi-LAT $\gamma$-Ray Pulsar Timing Array}",
      journal = {\apjl},
         year = 2024,
       volume = {963},
       number = {2},
          eid = {L46},
        pages = {L46},
          doi = {10.3847/2041-8213/ad2ae2},
archivePrefix = {arXiv},
       eprint = {2304.04735},
 primaryClass = {astro-ph.HE}
}

@article{Billard:2021uyg,
    author = "Billard, Julien and Boulay, Mark and Cebrian, Susana and Covi, Laura and Fiorillo, Giuliana and Green, Anne and Kopp, Joachim and Majorovits, Bela and Palladino, Kimberly and Petricca, Federica and Roszkowski, Leszek and Schumann, Marc",
    title = "{Direct Detection of Dark Matter -- APPEC Committee Report}",
    eprint = "2104.07634",
    archivePrefix = "arXiv",
    primaryClass = "hep-ex",
    doi = "10.1088/1361-6633/ac5754",
    journal = "Rept. Prog. Phys.",
    volume = "85",
    pages = "056201",
    year = "2022"
}

@article{Gaskins:2016cha,
    author = "Gaskins, Jennifer M.",
    title = "{A review of indirect searches for particle dark matter}",
    eprint = "1604.00014",
    archivePrefix = "arXiv",
    primaryClass = "astro-ph.HE",
    doi = "10.1080/00107514.2016.1175160",
    journal = "Contemp. Phys.",
    volume = "57",
    number = "4",
    pages = "496--525",
    year = "2016"
}

@article{DeRoeck:2024dm,
    author = "De Roeck, Albert",
    title = "{Dark matter searches at accelerators}",
    doi = "10.1016/j.nuclphysb.2024.116480",
    journal = "Nucl. Phys. B",
    volume = "1003",
    pages = "116480",
    year = "2024"
}

@article{LZ:2023First,
    author = "Aalbers, J. and others",
    collaboration = "LZ",
    title = "{First Dark Matter Search Results from the LUX-ZEPLIN (LZ) Experiment}",
    eprint = "2207.03764",
    archivePrefix = "arXiv",
    primaryClass = "hep-ex",
    doi = "10.1103/PhysRevLett.131.041002",
    journal = "Phys. Rev. Lett.",
    volume = "131",
    number = "4",
    pages = "041002",
    year = "2023"
}

@article{PandaX:2021First,
    author = "Meng, Yue and others",
    collaboration = "PandaX-4T",
    title = "{Dark Matter Search Results from the PandaX-4T Commissioning Run}",
    eprint = "2107.13438",
    archivePrefix = "arXiv",
    primaryClass = "hep-ex",
    doi = "10.1103/PhysRevLett.127.261802",
    journal = "Phys. Rev. Lett.",
    volume = "127",
    number = "26",
    pages = "261802",
    year = "2021"
}

@article{XENON:2023First,
    author = "Aprile, E. and others",
    collaboration = "XENON",
    title = "{First Dark Matter Search with Nuclear Recoils from the XENONnT Experiment}",
    eprint = "2303.14729",
    archivePrefix = "arXiv",
    primaryClass = "hep-ex",
    doi = "10.1103/PhysRevLett.131.041003",
    journal = "Phys. Rev. Lett.",
    volume = "131",
    number = "4",
    pages = "041003",
    year = "2023"
}

@article{FermiLAT:2015dSph,
    author = "Ackermann, M. and others",
    collaboration = "Fermi-LAT",
    title = "{Searching for Dark Matter Annihilation from Milky Way Dwarf Spheroidal Galaxies with Six Years of Fermi Large Area Telescope Data}",
    eprint = "1503.02641",
    archivePrefix = "arXiv",
    primaryClass = "astro-ph.HE",
    doi = "10.1103/PhysRevLett.115.231301",
    journal = "Phys. Rev. Lett.",
    volume = "115",
    number = "23",
    pages = "231301",
    year = "2015"
}

@article{HESS:2016Halo,
    author = "Abdallah, H. and others",
    collaboration = "H.E.S.S.",
    title = "{Search for Dark Matter Annihilations towards the Inner Galactic Halo from 10 Years of Observations with H.E.S.S.}",
    eprint = "1607.08142",
    archivePrefix = "arXiv",
    primaryClass = "astro-ph.HE",
    doi = "10.1103/PhysRevLett.117.111301",
    journal = "Phys. Rev. Lett.",
    volume = "117",
    number = "11",
    pages = "111301",
    year = "2016"
}

@article{Fan:2010gt,
    author = "Fan, JiJi and Reece, Matthew and Wang, Lian-Tao",
    title = "{Non-relativistic effective theory of dark matter direct detection}",
    eprint = "1008.1591",
    archivePrefix = "arXiv",
    primaryClass = "hep-ph",
    doi = "10.1088/1475-7516/2010/11/042",
    journal = "JCAP",
    volume = "11",
    pages = "042",
    year = "2010"
}

@article{Fitzpatrick:2012ix,
    author = "Fitzpatrick, A. Liam and Haxton, Wick and Katz, Emanuel and Lubbers, Nicholas and Xu, Yiming",
    title = "{The Effective Field Theory of Dark Matter Direct Detection}",
    eprint = "1203.3542",
    archivePrefix = "arXiv",
    primaryClass = "hep-ph",
    doi = "10.1088/1475-7516/2013/02/004",
    journal = "JCAP",
    volume = "02",
    pages = "004",
    year = "2013"
}

@article{XENON100:2017EFT,
    author = "Aprile, E. and others",
    collaboration = "XENON100",
    title = "{Effective field theory search for high-energy nuclear recoils using the XENON100 dark matter detector}",
    eprint = "1705.02614",
    archivePrefix = "arXiv",
    primaryClass = "astro-ph.CO",
    doi = "10.1103/PhysRevD.96.042004",
    journal = "Phys. Rev. D",
    volume = "96",
    number = "4",
    pages = "042004",
    year = "2017"
}

@article{XENON1T:2024EFT,
    author = "Aprile, E. and others",
    collaboration = "XENON",
    title = "{Effective field theory and inelastic dark matter results from XENON1T}",
    eprint = "2210.07591",
    archivePrefix = "arXiv",
    primaryClass = "hep-ex",
    doi = "10.1103/PhysRevD.109.112017",
    journal = "Phys. Rev. D",
    volume = "109",
    number = "11",
    pages = "112017",
    year = "2024"
}

@article{LZ:2024EFT,
    author = "Aalbers, J. and others",
    collaboration = "LZ",
    title = "{First constraints on WIMP-nucleon effective field theory couplings in an extended energy region from LUX-ZEPLIN}",
    eprint = "2312.02030",
    archivePrefix = "arXiv",
    primaryClass = "hep-ex",
    doi = "10.1103/PhysRevD.109.092003",
    journal = "Phys. Rev. D",
    volume = "109",
    number = "9",
    pages = "092003",
    year = "2024"
}

@article{TuckerSmith:2004jv,
    author = "Tucker-Smith, David and Weiner, Neal",
    title = "{The Status of Inelastic Dark Matter}",
    eprint = "hep-ph/0402065",
    archivePrefix = "arXiv",
    doi = "10.1103/PhysRevD.72.063509",
    journal = "Phys. Rev. D",
    volume = "72",
    pages = "063509",
    year = "2005"
}

@article{XENON100:2011iDM,
    author = "Aprile, E. and others",
    collaboration = "XENON100",
    title = "{Implications on Inelastic Dark Matter from 100 Live Days of XENON100 Data}",
    eprint = "1104.3121",
    archivePrefix = "arXiv",
    primaryClass = "astro-ph.CO",
    doi = "10.1103/PhysRevD.84.061101",
    journal = "Phys. Rev. D",
    volume = "84",
    pages = "061101",
    year = "2011"
}

@article{Baxter:2021pqo,
    author = "Baxter, D. and others",
    title = "{Recommended conventions for reporting results from direct dark matter searches}",
    eprint = "2105.00599",
    archivePrefix = "arXiv",
    primaryClass = "hep-ex",
    doi = "10.1140/epjc/s10052-021-09655-y",
    journal = "Eur. Phys. J. C",
    volume = "81",
    number = "10",
    pages = "907",
    year = "2021"
}

@article{McCabe:2013kea,
    author = "McCabe, Christopher",
    title = "{The Earth's velocity for direct detection experiments}",
    eprint = "1312.1355",
    archivePrefix = "arXiv",
    primaryClass = "astro-ph.CO",
    doi = "10.1088/1475-7516/2014/02/027",
    journal = "JCAP",
    volume = "02",
    pages = "027",
    year = "2014"
}

@article{Besla:2019lft,
    author = "Besla, Gurtina and Peter, Annika H. G. and Garavito-Camargo, Nicolas",
    title = "{The highest-speed local dark matter particles come from the Large Magellanic Cloud}",
    eprint = "1909.04140",
    archivePrefix = "arXiv",
    primaryClass = "astro-ph.GA",
    doi = "10.1088/1475-7516/2019/11/013",
    journal = "JCAP",
    volume = "11",
    pages = "013",
    year = "2019"
}

@article{SmithOrlik:2023lmc,
    author = "Smith-Orlik, Adam and Ronaghi, Nima and Bozorgnia, Nassim and Cautun, Marius and Fattahi, Azadeh and Besla, Gurtina and Frenk, Carlos S. and Garavito-Camargo, Nicolas and Gomez, Facundo A. and Grand, Robert J. J. and Marinacci, Federico and Peter, Annika H. G.",
    title = "{The impact of the Large Magellanic Cloud on dark matter direct detection signals}",
    eprint = "2302.04281",
    archivePrefix = "arXiv",
    primaryClass = "astro-ph.GA",
    doi = "10.1088/1475-7516/2023/10/070",
    journal = "JCAP",
    volume = "10",
    pages = "070",
    year = "2023"
}

@article{Arina:2014yna,
    author = "Arina, Chiara and Del Nobile, Eugenio and Panci, Paolo",
    title = "{Dark Matter with Pseudoscalar-Mediated Interactions Explains the DAMA Signal and the Galactic Center Excess}",
    eprint = "1406.5542",
    archivePrefix = "arXiv",
    primaryClass = "hep-ph",
    doi = "10.1103/PhysRevLett.114.011301",
    journal = "Phys. Rev. Lett.",
    volume = "114",
    pages = "011301",
    year = "2015"
}

@article{Dolan:2014ska,
    author = "Dolan, Matthew J. and Kahlhoefer, Felix and McCabe, Christopher and Schmidt-Hoberg, Kai",
    title = "{A taste of dark matter: Flavour constraints on pseudoscalar mediators}",
    eprint = "1412.5174",
    archivePrefix = "arXiv",
    primaryClass = "hep-ph",
    doi = "10.1007/JHEP03(2015)171",
    journal = "JHEP",
    volume = "03",
    pages = "171",
    year = "2015"
}

@article{Abe:2018emu,
    author = "Abe, Tomohiro and Fujiwara, Motoko and Hisano, Junji",
    title = "{Loop corrections to dark matter direct detection in a pseudoscalar mediator dark matter model}",
    eprint = "1810.01039",
    archivePrefix = "arXiv",
    primaryClass = "hep-ph",
    doi = "10.1007/JHEP02(2019)028",
    journal = "JHEP",
    volume = "02",
    pages = "028",
    year = "2019"
}

@article{Mimasu:2014nea,
    author = "Mimasu, Ken and Sanz, Veronica",
    title = "{ALPs at Colliders}",
    eprint = "1409.4792",
    archivePrefix = "arXiv",
    primaryClass = "hep-ph",
    doi = "10.1007/JHEP06(2015)173",
    journal = "JHEP",
    volume = "06",
    pages = "173",
    year = "2015"
}

@article{Dror:2023fyd,
    author = "Dror, Jeff A. and Gori, Stefania and Munbodh, Pankaj",
    title = "{QCD axion-mediated dark matter}",
    eprint = "2306.03145",
    archivePrefix = "arXiv",
    primaryClass = "hep-ph",
    doi = "10.1007/JHEP09(2023)128",
    journal = "JHEP",
    volume = "09",
    pages = "128",
    year = "2023"
}

\end{document}